\documentclass[aps, prd, reprint, floatfix, superscriptaddress, nofootinbib, showpacs, 10pt]{revtex4-1}

\usepackage{amsmath}
\usepackage{amssymb}
\usepackage{amsfonts}
\usepackage{amsthm}
\usepackage{mathtools}
\usepackage{mathrsfs}
\usepackage[caption=false, singlelinecheck=false,justification=raggedright]{subfig}
\usepackage{xspace}
\usepackage{txfonts}
\usepackage{multirow}
\usepackage{array}
\usepackage[normalem]{ulem}
\usepackage[colorlinks, urlcolor=black, citecolor=black, link color=black]{hyperref}
\newcommand{\modulus}[1]{\left| #1 \right|}

\def\nubar{\ensuremath{\overline{\nu}}}

\allowdisplaybreaks

\begin{document}

\title{Discovering intermediate mass sterile neutrinos through 
\texorpdfstring{$\tau^- \to \pi^- \mu^- e^+ \nu (\text{or }\overline{\nu})$}{tau- -> pi- mu- e+ nu} decay} %

\author{C.~S.~Kim} \email[E-mail at: ]{cskim@yonsei.ac.kr}%
\affiliation{Department of Physics and IPAP, Yonsei University, Seoul 120-749,
Korea} %

\author{G. L\'opez Castro}\email[E-mail at: ]{glopez@fis.cinvestav.mx}
\affiliation{Departamento de F\'isica, Centro de Investigaci\'on y de Estudios Avanzados,
Apartado Postal 14-740, 07000 M\'exico Distrito Federal, M\'exico}

\author{Dibyakrupa Sahoo} \email[E-mail at: ]{sahoodibya@yonsei.ac.kr}%
\affiliation{Department of Physics and IPAP, Yonsei University, Seoul 120-749,
Korea} %

\date{\today}

\begin{abstract}
Distinguishing the Dirac and Majorana nature of neutrinos remains one of the
most important tasks in neutrino physics. By assuming that the  $\tau^- \to
\pi^- \mu^- e^+ \nu (\text{or } \overline{\nu})$ decay is resonantly enhanced by
the exchange of an intermediate mass sterile neutrino $N$, we show that the
energy spectrum of emitted pions and muons can be used to easily distinguish
between the Dirac and Majorana nature of $N$. This method takes advantage of the
fact that the flavor of light neutrinos is not identified in the tau decay under
consideration. We find that it is particularly advantageous, because of no
competing background events, to search for $N$ in the mass range $m_e + m_{\mu}
\leqslant m_N \leqslant m_{\mu} + m_{\pi}$, where $m_X$ denotes the mass of
particle $X \in \{ e, \mu, \pi, N \}$.
\end{abstract}


\maketitle

\section{Introduction}

Lepton number is an absolutely conserved property of the standard model of
particle physics. However, observations of flavor oscillations in neutrinos have
revealed that neutrinos are massive as well as mixed with one another, opening
the interesting possibility that lepton number violation (LNV) occurs. Being
electrically neutral fermions, neutrinos can get their observed masses from the
well known Yukawa couplings (Dirac neutrinos) or from their self-conjugated
fields (Majorana neutrinos), in other words, they can be different or identical
to their own antiparticles, respectively. Another interesting consequence of
massive neutrinos is that the occurrence of lepton-flavor violation (LFV) in
decays of charged leptons is possible due to the mixing mechanism, although with
unobservably suppressed rates. Their observation at proposed flavor factories
experiments would indicate that mechanisms of LFV beyond the standard neutrino
mixing are necessary.

Among the important implications that neutrinos turn out to be Majorana
particles is that their masses originate beyond the Yukawa couplings via the
Higgs mechanism. The presence of Majorana mass terms for neutrinos entail  total
lepton number non-conservation  by two units ($\Delta L=2$), which would
manifest, for instance, in neutrinoless double-beta decays of nuclei, hadrons
and leptons. So far, $\Delta L=2$ violating processes have not been observed,
leading to strong constrains on their mass and mixing parameters
\cite{Rodejohann:2011mu}. In recent years, many studies have been reported on
the possibility of observing LNV by $\Delta L=2$ units in decays of mesons
\cite{mesons} and tau leptons \cite{tau} mediated by the exchange of a heavy
Majorana neutrino. The upper bounds on branching fractions obtained at flavor
factories \cite{tau-exp,mesons-exp} have allowed to set constraints on the
square of mixing of light ordinary neutrinos and heavy sterile neutrinos ranging
from $\modulus{V_{\ell N}}^2 =10^{-8} \sim 10^{-2}$ (where $\ell \in \{ e, \mu,
\tau \}$) as the neutrino mass increases from a few hundreds of MeV to a few
GeV.

The question of whether neutrinos are Dirac or Majorana particles remains one of
the most important questions for particle physics experiments. The existence of
heavy Majorana neutrinos would open the possibility to find mechanisms to
explain the smallness of active neutrino masses, as well as the viability of
theories to explain the baryon asymmetry or dark matter abundance in the
universe \cite{Asaka:2005an,Canetti:2014dka,Drewes:2015iva}. Different processes
that may be sensitive to the effects of light or heavy Majorana neutrinos have
been proposed and several experimental searches are underway. Beyond the
standard searches of nuclear neutrinoless double-beta decay experiments, several
studies have been suggested to distinguish between the Dirac and Majorana nature
of neutrinos, among others, precise measurements of neutrino-electron scattering
\cite{Kayser:1981nw,Barranco:2014cda}, the pseudo-Dalitz plot of sequential weak
decay processes involving $\nu\nubar$ pair as final products \cite{Kim:2016bxw},
and the spectrum of charged leptons from decays of intermediate on-shell heavy
neutrinos \cite{Cvetic:2015naa}.

The Dirac-type neutrinos can mediate processes with LFV but not LNV. Conversely,
Majorana neutrinos can intervene in processes with both LFV and LNV. In this
paper we use these properties of neutrinos to investigate how the study of
$\tau^-\to \pi^- \mu^- e^+\nu (\text{or } \nubar)$ decays can help to
disentangle the nature of the intermediate mass sterile neutrino mediating these
decays. Since the flavor of the light active neutrino in the final state is not
identified, it is not straightforward to distinguish scenarios with LFV ($\nu
=\nu_{e,\tau}$) from those with LNV ($\nubar=\nubar_{\mu}$). Among the four
channels possible in our case, the choice of a mono-energetic pion allows one to
single out two specific contributions. We show how, in the context of these two
chosen channels, the energy spectrum of the emitted muon can be useful to
distinguish between the Majorana and Dirac nature of the intermediate mass
sterile neutrino.

This paper is organized as follows: In Sec.~\ref{sec:partial-decay-width} we
compute the partial decay width of $\tau^- \to \pi^-\mu^-e^+ \nu (\text{or }
\nubar)$ decays assuming it to be facilitated by an on-shell intermediate mass
sterile neutrino. Here we discuss how by choosing mono-energetic pions we can
pick out the Feynman diagrams relevant to our methodology. Then in
Sec.~\ref{sec:muon-energy-spectrum} we discuss about the muon energy spectrum
and show how it helps in distinguishing between the Dirac and Majorana
possibilities for the intermediate mass neutrino. In Sec.~\ref{sec:comparison}
we compare our chosen modes with $\tau^- \to \mu^- \pi^+ \pi^-$ which has been
already searched for experimentally. Here we note that there exists a mass range
for $N$ in which we have no contamination from background events. Finally we
conclude in Sec.~\ref{sec:conclusion} reiterating the salient features of our
methodology.

\section{Partial decay width of the\\ \texorpdfstring{$\tau^- \to \pi^-\mu^-e^+ \nu (\text{or } \nubar)$}{tau- -> pi- mu- e+ nu} decays}\label{sec:partial-decay-width}

\begin{figure} 
\centering %
\subfloat[]{\label{fig:a}\includegraphics[width=0.4\linewidth,keepaspectratio]{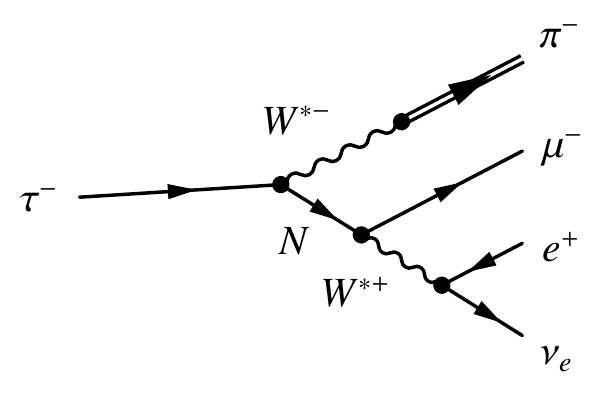} } \hfil%
\subfloat[]{\label{fig:b}\includegraphics[width=0.4\linewidth,keepaspectratio]{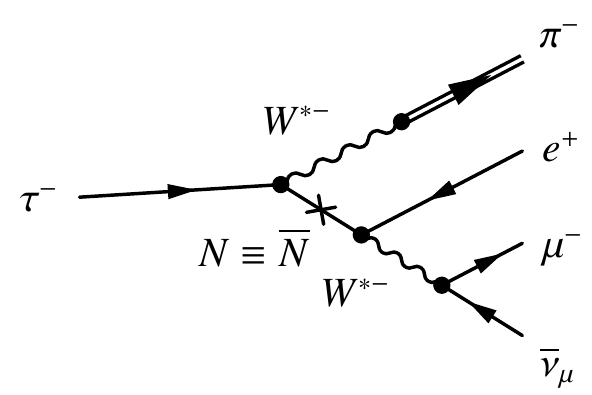} } \\%
\subfloat[]{\label{fig:c}\includegraphics[width=0.4\linewidth,keepaspectratio]{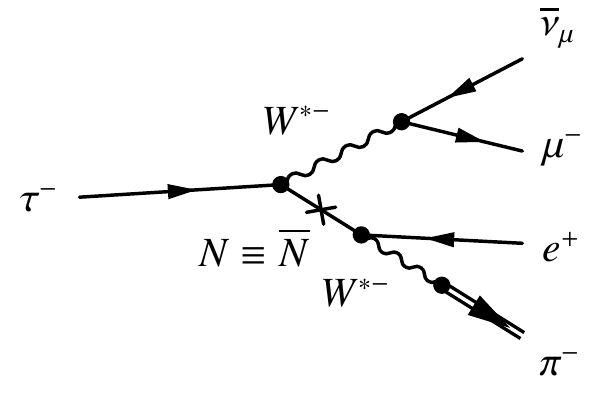} } \hfil%
\subfloat[]{\label{fig:d}\includegraphics[width=0.4\linewidth,keepaspectratio]{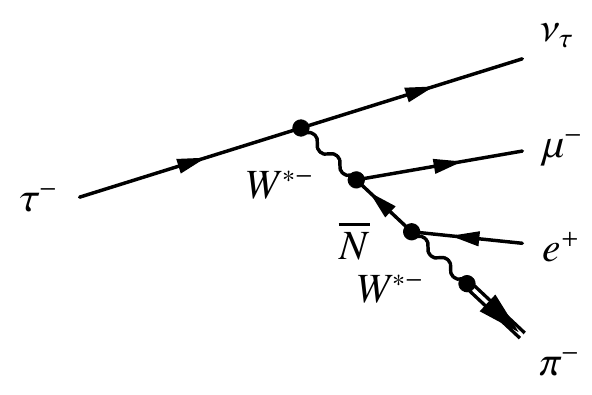} } %
\caption{Feynman diagrams contributions to $\tau^- \to \pi^- \mu^- e^+ \nu
(\text{or } \nubar)$, mediated by an intermediate mass sterile neutrino $N$ (the crossed
lines correspond to a Majorana particle). The light active neutrino (or
anti-neutrino) of flavor $\ell \in \{e,\mu,\tau\}$ allowed in each case is
denoted by $\nu_{\ell}$ (or $\nubar_{\ell}$).} %
\label{fig:figu1}
\end{figure}

Let us consider the $\tau^- \to \pi^-\mu^-e^+ \nu (\text{or } \nubar)$
decay\footnote{The analogous $\tau^- \to \pi^-\mu^+e^- \nu (\text{or } \nubar)$
decay follows a similar discussion under the exchange of $\mu \leftrightarrow e$
flavor labels.}, which we will assume to be mediated by the exchange of a single
on-shell (intermediate mass) sterile neutrino $N$, see Fig.~\ref{fig:figu1}. This decay
violates lepton flavor irrespective of the specific flavor identity of the
neutrino (or anti-neutrino) in the final state. If in the final state we have a
muon antineutrino ($\nubar_{\mu}$), the decay under consideration is a LNV
process and must be mediated by an intermediate mass Majorana neutrino $N$ as shown in
Figs.~\ref{fig:b} and \ref{fig:c}; otherwise, the intermediate neutrino $N$ can
be either Dirac or Majorana. Since, the flavor of the neutrino (or
anti-neutrino) in our final state can not be determined, it is not
straightforward to distinguish the Dirac and Majorana cases. Nevertheless, this
flavor ambiguity forces us to find some observables that can be used to
distinguish between the Dirac and Majorana neutrinos.

It is interesting to observe that the contributions in Figs.~\ref{fig:a} and
\ref{fig:b} can be differentiated from those in Figs.~\ref{fig:c} and
\ref{fig:d} by looking at the pion kinematics\footnote{Of course, if the
	lifetime of $N$ is large enough, those set of diagrams are also differentiated
	by the pion (muon) identification at the primary (secondary) or secondary
	(primary) vertices, respectively. Our analysis assumes no finite propagation of
	the intermediate mass neutrino with the displaced vertices.}. Owing to the
on-shell nature of the intermediate mass neutrino $N$, the pion energy spectrum
(in the rest frame of the tau lepton) is mono-energetic for Figs.~\ref{fig:a}
and \ref{fig:b} and has a continuous energy distribution for Figs.~\ref{fig:c}
and \ref{fig:d}. Conversely, the muon spectrum is continuous in all cases. In
the following we will focus in the case of a mono-energetic pion emission and
show how measurements of the branching ratio and muon energy spectrum in this
case allows us to distinguish between the effects of Dirac and Majorana
neutrinos.

Thus, let us consider only the Feynman diagrams shown in Figs.~\ref{fig:a} and
\ref{fig:b}. For given a mass $m_N$ of the sterile neutrino $N$, the distinctive
feature of Figs.~\ref{fig:a} and \ref{fig:b} is the mono-energetic pion with
energy $E_{\pi} = \left( m_{\tau}^2 + m_{\pi}^2 - m_N^2 \right) / 2m_{\tau}$.
The intermediate mass Majorana neutrino can be produced on-shell for mass values
in the range $m_e+m_{\mu} \leqslant m_N \leqslant m_{\tau}-m_{\pi}$, i.e.
$\approx 0.1061~\text{GeV} \leqslant m_N \leqslant 1.6372~\text{GeV}$. Using the
narrow-width approximation for $N$, the partial decay width of tau decay in our
case is given by:
\begin{equation}\label{dec-width}
\Gamma \left( \tau^- \to \pi^- \mu^- e^+ \nu \right) = \Gamma \left( \tau^- \to
\pi^- N \right) \frac{\Gamma \left( N \to \mu^- e^+ \nu (\text{or } \nubar)
\right)}{\Gamma_N},
\end{equation}
where $\Gamma_N$ is the full width of the sterile neutrino, and the partial decay
widths appearing in Eq.~\eqref{dec-width} are given, respectively, by:
\begin{widetext}
\begin{align}
\Gamma\left( \tau^- \to \pi^- N \right) &= \frac{G_F^2 f_{\pi}^2m_{\tau}^3
\modulus{V_{ud}}^2}{8\pi} \modulus{V_{\tau N}}^2\sqrt{\lambda\left( 1, r_N,
r_{\pi} \right)} \left[ \left( 1-r_N\right)^2-r_{\pi}\left(1+r_N\right)
\right],\label{TauNpi}\\%
\Gamma\left( N \to \mu^- e^+ \nu (\text{or } \nubar) \right) &= \Gamma\left( N
\to \mu^- e^+ \nu_e\right)+\Gamma\left(N \to \mu^- e^+ \nubar_{\mu} \right) =
\frac{G_F^2 m_N^5\modulus{V_{\mu N}}^2}{192\pi^3 } \left(
1-8r_{\mu}+8r_{\mu}^3-r_{\mu}^4-12r_{\mu}^2\ln r_{\mu}\right) \left( 1 + \alpha
~ R_{e\mu} \right),\label{eq:N2MuENu-DM}\\ %
&=
\begin{cases}
\displaystyle \frac{G_F^2 m_N^5\modulus{V_{\mu N}}^2}{192\pi^3 } \left(
1-8r_{\mu}+8r_{\mu}^3-r_{\mu}^4-12r_{\mu}^2\ln r_{\mu}\right), & 
\text{(if $N$ is a Dirac neutrino, $\alpha=0$)}\\[3mm]%
\displaystyle \frac{G_F^2 m_N^5\modulus{V_{\mu N}}^2}{192\pi^3 } \left(
1-8r_{\mu}+8r_{\mu}^3-r_{\mu}^4-12r_{\mu}^2\ln r_{\mu}\right) \left( 1 +
R_{e\mu} \right). & \text{(if $N$ is a Majorana neutrino, $\alpha = 1$)}
\end{cases}\label{eq:N2MuENu}
\end{align}
\end{widetext}
where $f_{\pi} = 130.2$~MeV \cite{PDG2016}, $r_{N,\pi}=m^2_{N,\pi}/m^2_{\tau}$,
$r_{\mu}=m_{\mu}^2/m_N^2$, with $\modulus{V_{\ell N}}^2$ denoting the mixing of
active neutrino of flavor $\ell = e,\mu,\tau$ with the sterile neutrino $N$. In
the above expressions, $\alpha$ is a parameter that allows to distinguish the
sterile intermediate Dirac ($\alpha=0$) and Majorana ($\alpha=1$)
neutrinos\footnote{Note that a Majorana neutrino ($\alpha =1$) with $R_{e\mu}
	\to 0$ can not be distinguished here from a Dirac neutrino ($\alpha =0$).
	However, if we instead consider the decay $\tau^- \to \pi^- \mu^+ e^- \nu
	(\text{or } \nubar)$ and look at the pion and muon energy distributions as
	discussed in this paper, the above problem does not appear.}, $R_{e\mu} =
\modulus{V_{eN}}^2/\modulus{V_{\mu N}}^2$, and $\lambda\left(x,y,z\right) = x^2
+ y^2 + z^2 - 2 \left(xy+yz+zx\right)$ is the K\"all\'en function. It is easy to
check that Eq.~\eqref{TauNpi} gives the usual expression for the decay width of
$\tau^-\to \pi^-\nu_{\tau}$ when one takes $r_N\to 0$ and $\modulus{V_{\tau
		N}}^2\to 1$. Similarly, when $\alpha=0$ and $\modulus{V_{\ell N}}^2=1$ (the case
of Dirac neutrino) Eq.~\eqref{eq:N2MuENu} becomes identical to the well known
rate of muon decay in the crossed channel.

In order to provide an estimate of the $\tau^- \to \pi^- \mu^- e^+ \nu (\text{or
} \nubar)$ branching fraction we need an input for the total width $\Gamma_N$.
From the sum over all the exclusive decay channels that open below $m_N$
\cite{Cvetic:2014nla} we get the following expressions for total decay width of
$N$ for two typical values of the neutrino mass:
\begin{align}
\Gamma_N &\approx \frac{G_F^2 m_N^5}{96 \pi^3} \left( 15 \modulus{V_{eN}}^2   +
8 \modulus{V_{\mu N}}^2 + 2 \modulus{V_{\tau N}}^2 \right), && \left(
\substack{\text{for}\\[2mm] m_N = 0.25~\textrm{GeV}} \right)\\ %
\Gamma_N &\approx \frac{G_F^2 m_N^5}{96 \pi^3} \left( 7 \modulus{V_{eN}}^2 + 7
\modulus{V_{\mu N}}^2 + 2 \modulus{V_{\tau N}}^2 \right). && \left(
\substack{\text{for}\\[2mm]m_N = 1~\textrm{GeV}} \right)
\end{align}
Using the upper bounds for the mixing elements $\modulus{V_{\ell N}}^2$ that
were reported in Ref.~\cite{Cvetic:2014nla}, we evaluate the decay width
$\Gamma_N$ of the intermediate mass neutrino. The results are given in
Table~\ref{tab1}.

\begin{table}[hbtp]
\centering
\begin{tabular}{|c|c|c|c|l|} \hline % 
$m_N$ (in GeV) & $\modulus{V_{eN}}^2$ & $\modulus{V_{\mu N}}^2$ &
$\modulus{V_{\tau N}}^2$ & $\Gamma_N$ (in GeV) \\ \hline %
$0.25$ & $10^{-8}$ & $10^{-7}$ & $10^{-4}$ & $8.97 \times 10^{-21}$\\ \hline %
$1.0$ & $10^{-7}$ & $10^{-7}$ & $10^{-2}$ & $9.14 \times 10^{-16}$\\ \hline 
\end{tabular}
\caption{Upper bounds on the mixings of sterile neutrinos with light active
	neutrinos \cite{Cvetic:2014nla} and total decay width of the sterile neutrino
	for two reference values of $m_N$.}%
\label{tab1}
\end{table}

Finally, the branching fraction for the $\tau^- \to \pi^- \mu^- e^+ \nu
(\text{or } \nubar)$ decay can be obtained by dividing Eq. (\ref{dec-width}) by
the measured width of the $\tau$ lepton \cite{PDG2016}. The estimated upper
limits for the partial widths involved Eqs.~\eqref{TauNpi} and
\eqref{eq:N2MuENu} are shown in Table \ref{tab2} for the two values of $m_N$,
and the branching fraction for the $\tau^- \to \pi^- \mu^- e^+ \nu (\text{or }
\nubar)$ decay are shown in the last column.

\begin{table}[hbtp]
\centering
\begin{tabular}{|c|c|c|l|}\hline
$m_N$ & $\Gamma \left( \tau^- \to \pi^- N \right)$ & $\Gamma \left( N \to \mu^-
e^+ \nu (\text{or } \nubar) \right)$ & {\scriptsize Branching fraction for} \\ %
{\scriptsize(in GeV)} & {\scriptsize(in GeV)} & {\scriptsize(in GeV)} & $\tau^-
\to \pi^- \mu^- e^+ \nu (\text{or } \nubar)$ \\ \hline %
$0.25$ & $4.86 \times 10^{-17}$ & $6.14\times 10^{-25} \Big( 1 +
\frac{\alpha}{10} \Big)$ & $1.47 \times 10^{-9} \Big( 1 + \frac{\alpha}{10}
\Big)$ \\ \hline %
$1$ & $1.61\times 10^{-15}$ & $2.10 \times 10^{-21} \Big( 1 + \alpha \Big)$ &
$1.63\times10^{-9} \Big( 1 + \alpha \Big)$ \\ \hline
\end{tabular}
\caption{Upper bounds on the partial decay widths for production and decay of
	the intermediate mass sterile neutrino $N$, and the branching fraction for the
	$\tau^- \to \pi^- \mu^- e^- \nu$ decay.} %
\label{tab2}
\end{table}

We conclude that, with current bounds on the mixing elements $\modulus{V_{\ell
N}}^2$, the effects  the branching fraction of the tau decay under consideration
may be more sensitive to the effects of Majorana neutrinos for larger values of
heavy neutrino mass. Those branching fractions lie at the edge of  Belle~II
capabilities, which is expected to produce about $\mathcal{O}(10^{10})$ tau
lepton pairs in the full dataset \cite{Inami:2016aba}. Conversely, from an
experimental upper limit on the $\tau^- \to \pi^- \mu^- e^+ \nu (\text{or }
\nubar)$ decay at future flavor factories, one can constrain the parameter space
$\left( \modulus{V_{\ell N}}^2, m_N \right)$ under the assumption that the
exchanged sterile neutrino is a Dirac or Majorana particle.

\section{Muon energy spectrum in the \texorpdfstring{$\tau^- \to \pi^- \mu^- e^+ \nu (\text{or } \nubar)$}{tau- -> pi- mu- e+ nu} decays}\label{sec:muon-energy-spectrum}

From Eq.~\eqref{eq:N2MuENu} it is clear that the decay rate for $\tau^- \to
\pi^- \mu^- e^+ \nu (\text{or } \nubar)$ (more precisely, the decay rate of the
intermediate mass neutrino $N$) is the same for both Dirac and Majorana cases up
to an overall normalization factor that depends on the nature of the exchanged
neutrino. Therefore, in order to distinguish between the Majorana and Dirac
cases, we have to look at another observable like the muon energy spectrum.

\begin{figure}[hbtp]
\centering %
\subfloat[]{\includegraphics[width=\linewidth,
keepaspectratio]{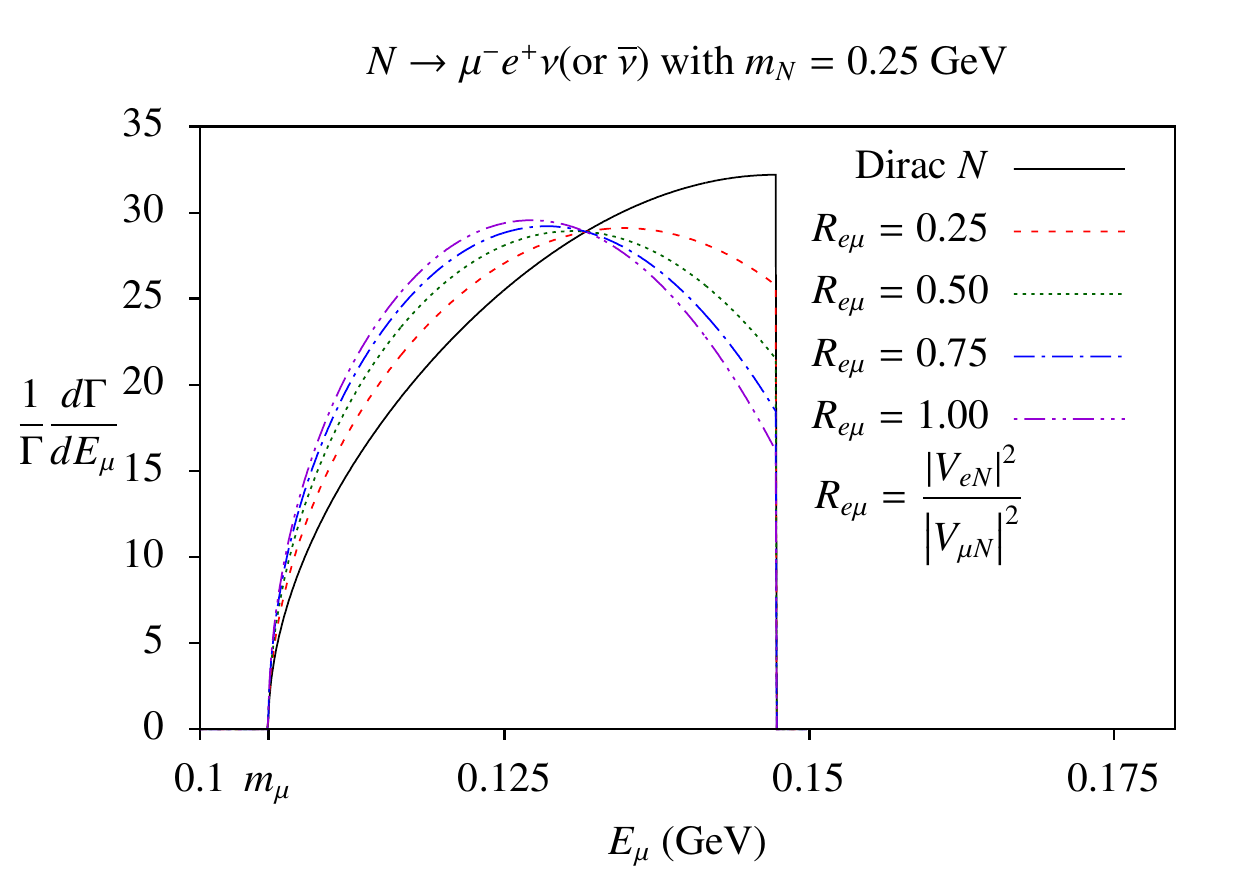}} \\ %
\subfloat[]{\includegraphics[width=\linewidth,
keepaspectratio]{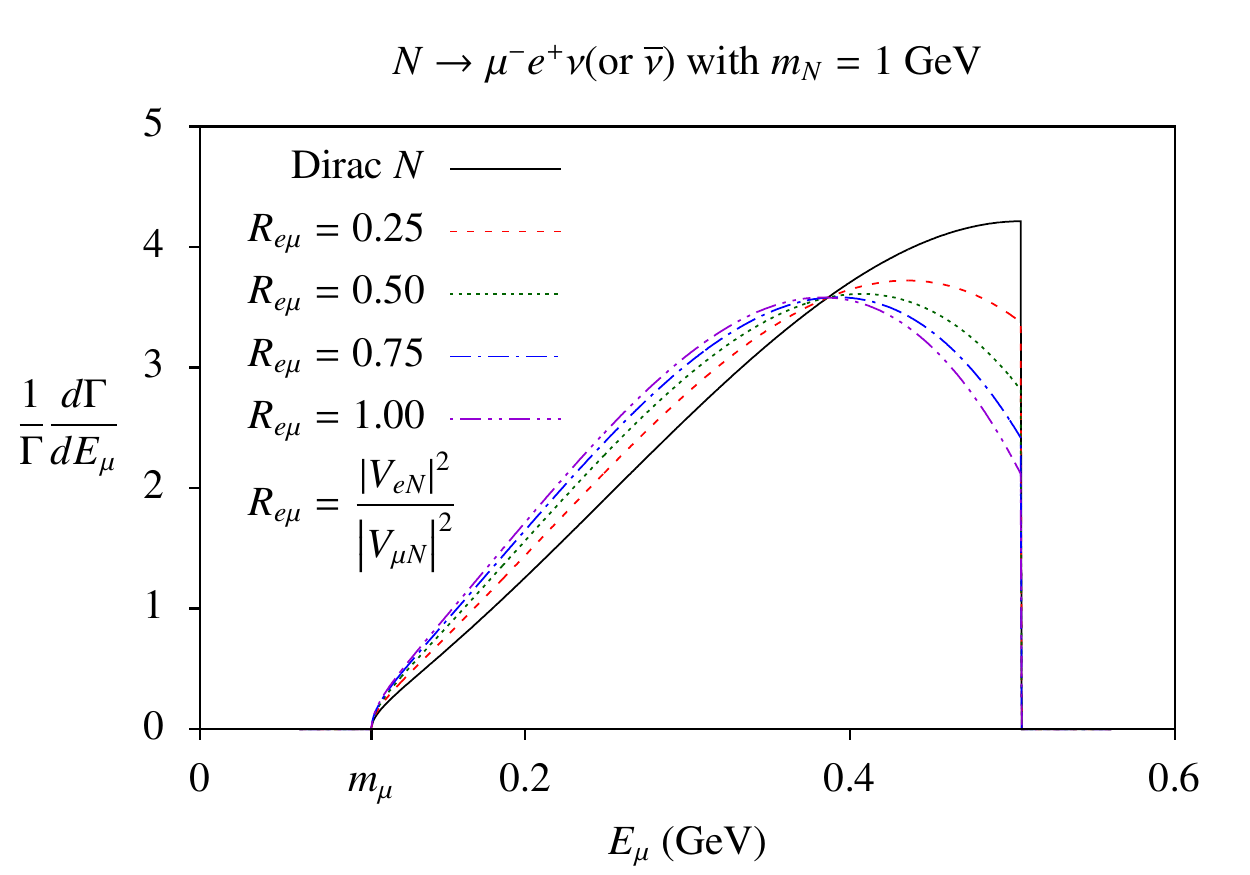}} %
\caption{Normalized muon energy distribution for the decay $N \to \mu^- e^+
\nu(\textrm{or}~\overline{\nu})$ where the flavor of the final neutrino (or
anti-neutrino) is unknown. The upper (lower) plot corresponds to $m_N=0.25$ GeV
($m_N=1$ GeV). The solid curves correspond to the case with Dirac neutrino
($\alpha=0$), while the other curves correspond to the case with Majorana
neutrino ($\alpha =1$) for different choices of $R_{e\mu}$. It should again be
noted that if the Majorana neutrino $N$ has vanishingly small mixing with
electron-type neutrino, i.e.\ $R_{e\mu} \to 0$, then such a neutrino can not be
distinguished from the Dirac neutrino case using the plots shown here.
Nevertheless, this special case can be easily addressed if we consider the
$\tau^- \to \pi^-\mu^+ e^- \nu (\text{or } \nubar)$ decays and follow our
methodology of looking at the pion and muon energies as discussed in the main
text. All experimental values were taken from PDG 2016~\cite{PDG2016}.} %
\label{fig:N2muenu-general}
\end{figure}

The normalized muon energy distribution for the decay $N \to \mu^- e^+ \nu
(\text{or } \nubar)$ in the rest frame of $N$ is given by,
\begin{widetext}
\begin{equation}
\frac{1}{\Gamma\left( N \to \mu^- e^+ \nu (\text{or } \nubar) \right)}
\frac{d\Gamma\left( N \to \mu^- e^+ \nu (\text{or } \nubar) \right)}{dE_{\mu}} =
\frac{\displaystyle 96 m_N^3 \left( - \frac{1}{3} m_N m_{\mu}^2 + \left(
\frac{1}{2} + \alpha R_{e\mu} \right) \left( m_N^2 + m_{\mu}^2 \right) E_{\mu} -
\left(\frac{2}{3} + 2 \alpha R_{e\mu} \right) m_N E_{\mu}^2 \right)
\sqrt{E_{\mu}^2 - m_{\mu}^2}}{\displaystyle \left( \left( m_N^4 - m_{\mu}^4
\right) \left( m_N^4 + m_{\mu}^4 - 8 m_{\mu}^2 m_N^2 \right) - 24 m_{\mu}^4
m_N^4 \log\left(\frac{m_{\mu}}{m_N}\right)\right) \left( 1 + \alpha R_{e\mu}
\right)}.
\end{equation}
\end{widetext}
The muon energy distributions for $m_N=0.25$~GeV and $1$~GeV and for various
values of $R_{e\mu}$ are shown in Fig.~\ref{fig:N2muenu-general}. It is clear
from these plots that the shape of the muon spectrum can help to differentiate
between the contributions of the intermediate mass Dirac and Majorana neutrinos.
A peak of the muon energy spectrum at the end of the maximum allowed muon energy
would signal that the intermediate mass neutrino is a Dirac particle.
Conversely, a peaked spectrum towards lower energies (i.e.\ away from the
kinematic end-point of $E_{\mu}$) would indicate its Majorana nature.

\section{Comparison with \texorpdfstring{$\tau^- \to \mu^- \pi^+ \pi^-$}{tau- -> mu- pi+ pi-} mode, and discussion on some possible new physics contributions}\label{sec:comparison}

\begin{figure}[hbtp]
\centering%
\includegraphics[width=\linewidth,keepaspectratio]{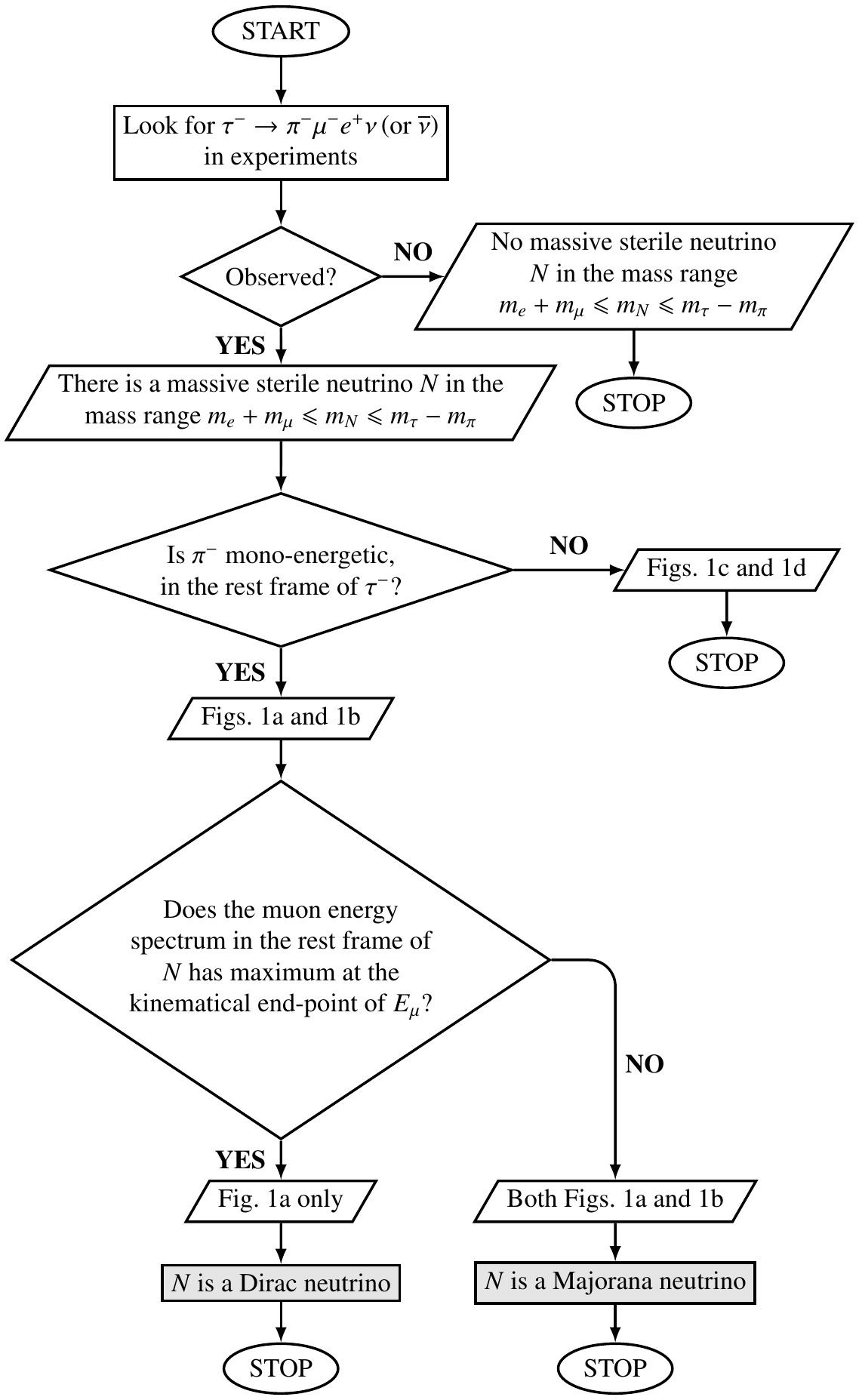}%
\caption{The flow chart for disentangling the Dirac and Majorana nature of the
	exchanged intermediate mass neutrino $N$ in the decay $\tau^- \to \pi^- \mu^-
	e^+ \nu(\textrm{or}~\overline{\nu})$. This flowchart must be read in conjunction
	with Fig.~\ref{fig:figu1}.} %
\label{fig:flowchart}
\end{figure}

\begin{figure}[hbtp]
\centering%
\includegraphics[width=\linewidth]{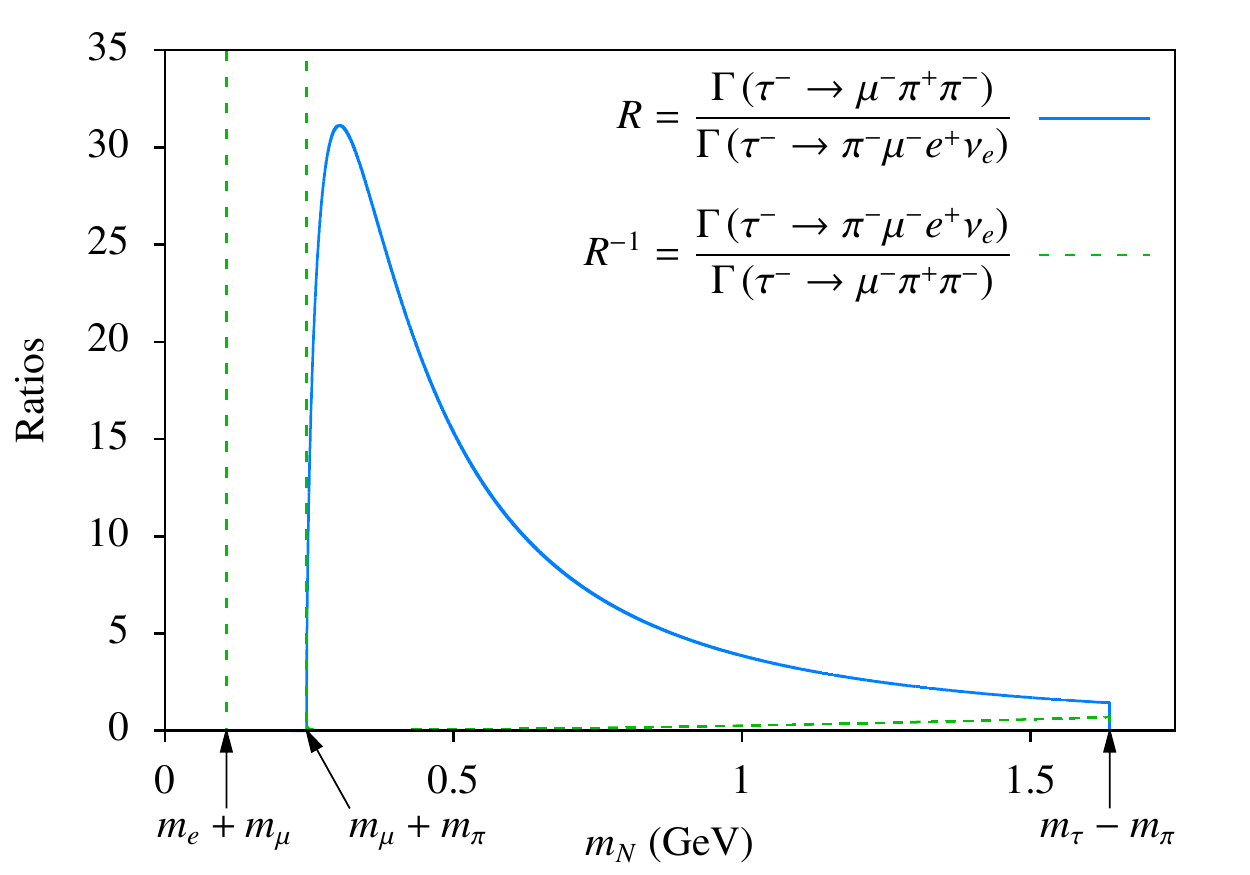}%
\caption{Comparison between the decay rates of $\tau^- \to \mu^- \pi^+ \pi^-$
and $\tau^- \to \pi^-\mu^- e^+ \nu_e$ assuming that both are mediated by an intermediate mass
on-shell sterile neutrino of mass $m_N$, with $m_{\mu} + m_{\pi} \leqslant m_N
\leqslant m_{\tau}-m_{\pi}$. Here we have also used the narrow-width
approximation for $N$. In the mass range $m_e + m_{\mu} \leqslant m_N \leqslant
m_{\mu} + m_{\pi}$ the $\tau^- \to \pi^- \mu^- e^+ \nu_e$ is allowed, but
$\tau^- \to \mu^- \pi^+ \pi^-$ is not. Therefore, in this mass range there are
no background events for the decay mode $\tau^- \to \pi^-\mu^- e^+ \nu_e$. The
dashed line in the mass region $m_e + m_{\mu} \leqslant m_N \leqslant m_{\mu} +
m_{\pi}$ rises to infinity.}%
\label{fig:comparison}
\end{figure}

Fig.~\ref{fig:flowchart} summarizes the procedure that needs to be followed to
establish the nature of the intermediate mass sterile neutrino exchanged in
$\tau^- \to \pi^-\mu^-e^+ \nu (\text{or } \nubar)$ decays (here flavor of the
final light neutrino or anti-neutrino is unknown) with anticipation that this
three-prong decay is observed at future colliders, such as Belle~II. In this
context it is important to note that the lepton flavor violating decay $\tau^-
\to \mu^- \pi^+ \pi^-$ can also contribute to our decay channels via the
sequential decay $\pi^+ \to e^+ \nu_e$. Such lepton flavor violating decays have
already been searched for in experiments~\cite{Aubert:2005tp, Miyazaki:2012mx}.
To make a quantitative comparison between the decay modes $\tau^- \to \mu^-
\pi^+ \pi^-$ and $\tau^- \to \pi^-\mu^-e^+ \nu_e$ we define the ratio, $R =
\Gamma\left( \tau^- \to \mu^- \pi^+ \pi^- \right) / \Gamma\left( \tau^- \to
\pi^-\mu^-e^+ \nu_e \right)$. Assuming that both the decay modes are facilitated
by the exchange of an intermediate mass neutrino and applying the narrow-width
approximation for it, we get the result as shown in Fig.~\ref{fig:comparison}.
Note that in both the decay modes we consider the $\pi^-$ to be mono-energetic
in the rest frame of $\tau^-$. From Fig.~\ref{fig:comparison} it is very clear
that for lower values of $m_N$ the $\tau^- \to \mu^- \pi^+ \pi^-$ has much
larger branching ratio than $\tau^- \to \pi^-\mu^- e^+ \nu_e$. In the region
$m_{e} + m_{\mu} \leqslant m_N \leqslant m_{\mu} + m_{\pi}$ there is no
contribution from the $\tau^- \to \mu^- \pi^+ \pi^-$ mode due to phase space
considerations, however, $\tau^- \to \pi^- \mu^- e^+ \nu(\text{or}~\nubar)$ can
still give contributions as long as $m_N > m_e + m_{\mu}$ (see the dashed line
in Fig.~\ref{fig:comparison} which goes to infinity in the mass region $m_{e} +
m_{\mu} \leqslant m_N \leqslant m_{\mu} + m_{\pi}$). Due to lack of any
background events in the mass range $m_{e} + m_{\mu} \leqslant m_N < m_{\mu} +
m_{\pi}$, it is the experimentally clean region to study the decay mode $\tau^-
\to \pi^-\mu^- e^+ \nu_e$. It is thus expected that observing the decay modes
$\tau^- \to \pi^-\mu^-e^+ \nu (\text{or } \nubar)$ ought to be feasible with
future colliders and following the flowchart of Fig.~\ref{fig:flowchart} it
would be possible to decipher the nature of the heavy intermediate on-shell
sterile neutrino.

\begin{figure}[hbtp]
\centering%
\includegraphics[width=\linewidth,keepaspectratio]{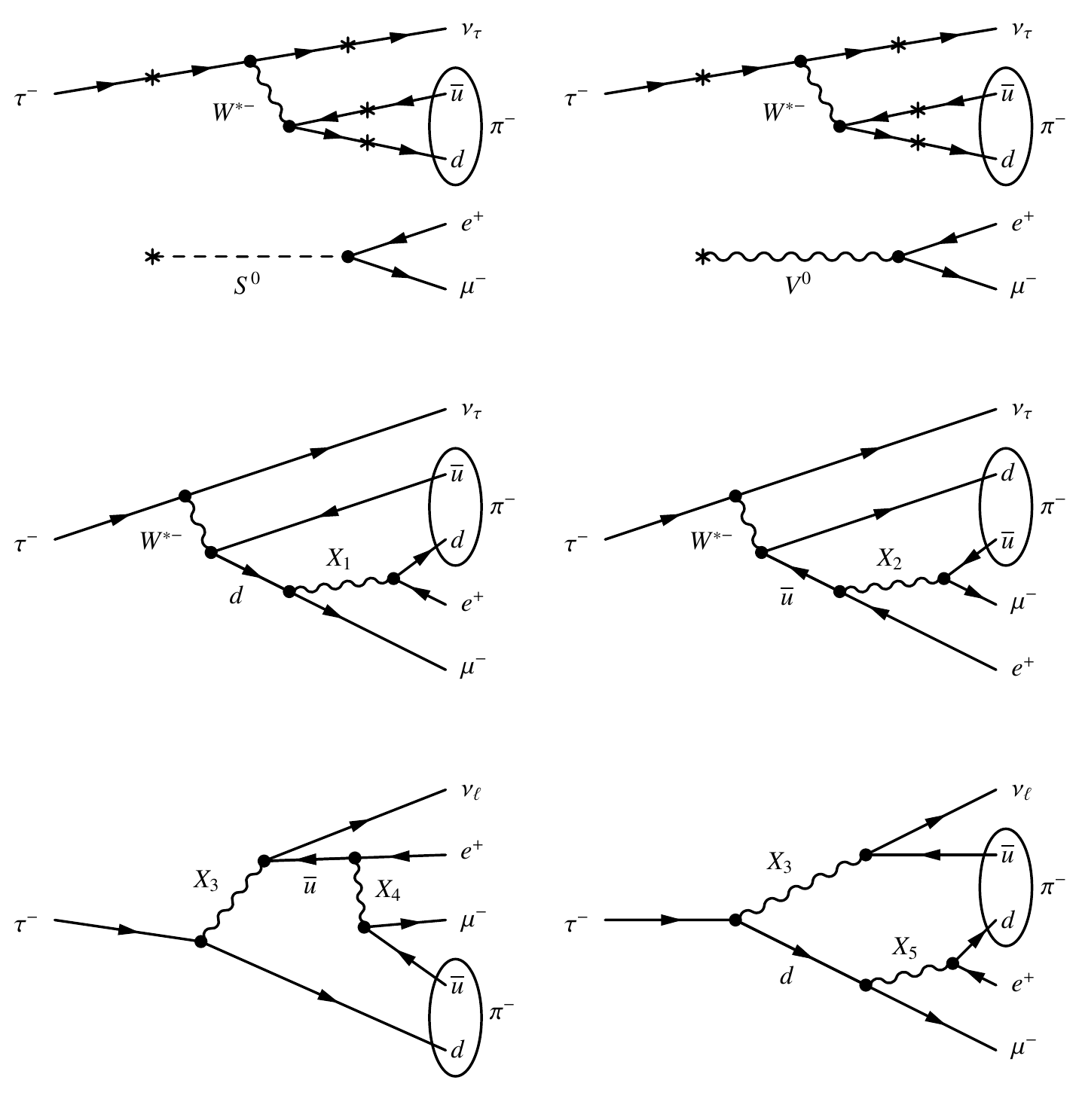}
\caption{Some generic new physics contributions to $\tau^- \to \pi^- \mu^- e^+
	\nu$ (or $\overline{\nu}$) other than the intermediate mass sterile neutrino
	$N$. The new physics contributions can arise from some scalars ($S^0$) or
	vectors ($V^0$) or lepto-quarks ($X_i$, with $i=1,\ldots,5$). Here the asterisk
	($*$) denotes the four possibilities of connecting the $S^0$ and $V^0$ lines to
	the four fermion lines of the accompanying diagram.}%
\label{fig:NP}
\end{figure}

Note that there can be some other generic new physics contributions, other than
the intermediate mass sterile neutrinos, to our decay mode $\tau^- \to \pi^-
\mu^- e^+ \nu$ (or $\overline{\nu}$). Some such possibilities are shown in
Fig.~\ref{fig:NP}. As is clear from Fig.~\ref{fig:NP} there can be some
lepton-flavor violating new physics via some scalar ($S^0$) or vector ($V^0$)
particle (e.g. lepton flavor violating modes of Higgs, or $Z$, or some $Z'$), as
well as some lepto-quark ($X_i$, with $i=1,\ldots, 5$) contributions. It must be
noted that in all these new physics possibilities, the pion is \textsl{not}
mono-energetic. Hence, following the flowchart of Fig.~\ref{fig:flowchart}, we
can avoid these new physics possibilities completely. Furthermore, the lepton
flavor violating modes of Higgs, $Z$, as well as those of $Z'$ are severely
constrained by experimental data. The lepto-quarks $X_1, \ldots, X_5$ in
Figs.~\ref{fig:NP} also facilitate lepton-flavor violation, and must therefore
be severely constrained by $\mu \to e \gamma$ searches. Thus, other new physics
possibilities for $\tau^- \to \pi^- \mu^- e^+ \nu$ (or $\overline{\nu}$) decay
are not only very constrained by existing data, but their presence, if any, does
not affect our analysis as they are easily discarded by considering
mono-energetic pions as emphasized in our methodology here.

\section{Conclusion}\label{sec:conclusion}

From the four possible contributions to the decay $\tau^- \to \pi^-\mu^-e^+ \nu
(\text{or } \nubar)$ as shown in Fig.~\ref{fig:figu1}, choosing a mono-energetic
$\pi^-$ in the rest frame of $\tau^-$ should allow us to isolate two of them
(Figs.~\ref{fig:a} and \ref{fig:b}). The spectrum of muons produced from the
decay of the intermediate mass neutrino $N$ in Figs.~\ref{fig:a} and
\ref{fig:b}, would indicate that a peak observed in the muon energy spectrum
below its kinematic endpoint corresponds to an intermediate mass Majorana
neutrino. Thus by a clever analysis of the pion and muon energy spectra, we can
easily distinguish between the Dirac and Majorana possibilities for the
intermediate on-shell sterile neutrino.

\textbf{Note added:} A related idea to the one presented in this paper was
reported in Ref.~\cite{Yuan:2017xdp}. Our work differs in several ways. First,
contrary to us, Ref.~\cite{Yuan:2017xdp} considers the decay $\tau^+ \to
\pi^-e^+e^+\nu$, where the $\tau$ lepton and $\pi$ meson have opposite charges
(or same-sign charged leptons in the final state) which requires that two
separate vertices are detected to avoid the exchange of identical fermions.
Second, the authors of Ref.~\cite{Yuan:2017xdp} compute the pion instead of the
muon spectrum.  This procedure, however, will not allow to distinguish the
nature of the exchanged neutrino since the pion spectra for Dirac and Majorana
cases differ only in normalization but not in shape.

\acknowledgments 

The work of C.S.K. was supported by the NRF grant funded by the Korean
government of the MEST (No.~2016R1D1A1A02936965). G.L.C. is grateful to Conacyt
for financial support under Projects No.~236394, No.~250628 (Ciencia
B\'{a}sica), and No.~296 (Fronteras de la Ciencia). C.S.K and D.S. would like to
thank CINVESTAV, Mexico, where some initial part of this work was done, for
hospitality.


\begin{thebibliography}{99}
%
\bibitem{Rodejohann:2011mu} 
  W.~Rodejohann,
  %``Neutrino-less Double Beta Decay and Particle Physics,''
  Int.\ J.\ Mod.\ Phys.\ E {\bf 20}, 1833 (2011)
%
\bibitem{mesons}
C.~Dib, V.~Gribanov, S.~Kovalenko and I.~Schmidt,
  %``K meson neutrinoless double muon decay as a probe of neutrino masses and mixings,''
  Phys.\ Lett.\ B {\bf 493}, 82 (2000);
C.~Dib, V.~Gribanov, S.~Kovalenko and I.~Schmidt,
  %``Lepton number violating processes and Majorana neutrinos,''
  Part.\ Nucl.\ Lett.\  {\bf 106}, 42 (2001); 
D.~Delepine, G.~Lopez Castro and N.~Quintero,
  %``Lepton number violation in top quark and neutral B meson decays,''
  Phys.\ Rev.\ D {\bf 84}, 096011 (2011),
  Erratum: [Phys.\ Rev.\ D {\bf 86}, 079905(E) (2012)]; 
G.~Lopez Castro and N.~Quintero,
  %``Bounding resonant Majorana neutrinos from four-body B and D decays,''
  Phys.\ Rev.\ D {\bf 87}, 077901 (2013); 
D.~Milanes, N.~Quintero and C.~E.~Vera,
  %``Sensitivity to Majorana neutrinos in $\Delta L=2$ decays of $B_c$ meson at LHCb,''
  Phys.\ Rev.\ D {\bf 93}, no. 9, 094026 (2016);
H.~Yuan, T.~Wang, G.~L.~Wang, W.~L.~Ju and J.~M.~Zhang,
  %``Lepton-number violating four-body decays of heavy mesons,''
  JHEP {\bf 1308}, 066 (2013);
G.~Cvetic, C.~Dib, S.~K.~Kang and C.~S.~Kim,
  %``Probing Majorana neutrinos in rare K and D, ~D_s, B, B_c meson decays,''
  Phys.\ Rev.\ D {\bf 82}, 053010 (2010);
G.~Cvetic, C.~Dib and C.~S.~Kim,
  %``Probing Majorana neutrinos in rare $\pi^+ \to e^+ e^+ \mu^- \nu$ decays,''
  JHEP {\bf 1206}, 149 (2012);
C.~Dib and C.~S.~Kim,
  %``Remarks on the lifetime of sterile neutrinos and the effect on detection of rare meson decays $M^+ \to M^{\prime}-\ell^+\ell^+$,''
  Phys.\ Rev.\ D {\bf 89}, no. 7, 077301 (2014);
G.~Cvetic, C.~S.~Kim, Y.~J.~Kwon and Y.~M.~Yook,
  %``Decay of $B^\pm \to \tau^\pm +$ "missing momentum" and direct measurement of the mixing parameter $U_{\tau N}$,''
  Phys.\ Rev.\ D {\bf 93}, no. 1, 013003 (2016);
G.~Cvetic and C.~S.~Kim,
  %``Rare decays of B mesons via on-shell sterile neutrinos,''
  Phys.\ Rev.\ D {\bf 94}, no. 5, 053001 (2016),
  Erratum: [Phys.\ Rev.\ D {\bf 95}, no. 3, 039901(E) (2017)];
G.~Cvetic, F.~Halzen, C.~S.~Kim and S.~Oh,
  %``Anomalies in (Semi)-Leptonic $B$ Decays $B^{\pm} \to \tau^{\pm} \nu$, $B^{\pm} \to D \tau^{\pm} \nu$ and $B^{\pm} \to D^* \tau^{\pm} \nu$, and Possible Resolution with Sterile Neutrino,''
  arXiv:1702.04335 [hep-ph];
G.~Cvetic and C.~S.~Kim,
  %``Sensitivity limits on heavy-light mixing $|U_{\mu N}|^2$ from lepton number violating $B$ meson decays,''
  arXiv:1705.09403 [hep-ph];
S.~Mandal and N.~Sinha,
  %``Favoured $B_c$ Decay modes to search for a Majorana neutrino,''
  Phys.\ Rev.\ D {\bf 94}, no. 3, 033001 (2016);
B.~Shuve and M.~E.~Peskin,
  %``Revision of the LHCb Limit on Majorana Neutrinos,''
  Phys.\ Rev.\ D {\bf 94}, no. 11, 113007 (2016);
C.~O.~Dib, C.~S.~Kim and K.~Wang,
  %``Search for Heavy Sterile Neutrinos in Trileptons at the LHC,''
  arXiv:1703.01936 [hep-ph] (to be published in Chinese Physics C.);
C.~O.~Dib, C.~S.~Kim and K.~Wang,
  %``Signatures of Dirac and Majorana sterile neutrinos in trilepton events at the LHC,''
  Phys.\ Rev.\ D {\bf 95}, no. 11, 115020 (2017).

%%

\bibitem{tau}
C.~Dib, J.~C.~Helo, M.~Hirsch, S.~Kovalenko and I.~Schmidt,
  %``Heavy Sterile Neutrinos in Tau Decays and the MiniBooNE Anomaly,''
  Phys.\ Rev.\ D {\bf 85}, 011301 (2012);
G.~L.~Castro and N.~Quintero,
  %``Lepton number violating four-body tau lepton decays,''
  Phys.\ Rev.\ D {\bf 85}, 076006 (2012),
  Erratum: [Phys.\ Rev.\ D {\bf 86}, 079904(E) (2012)]; 
D.~Gomez Dumm and P.~Roig,
  %``Dispersive analysis of $\tau^-\to\pi^-\pi^0\nu_\tau$ Belle data,''
  Nucl.\ Phys.\ Proc.\ Suppl.\  {\bf 253-255}, 12 (2014);
J.~Zamora-Saa,
%``Resonant $CP$ violation in rare $\tau^{\pm}$ decays,''
JHEP {\bf 1705}, 110 (2017);
G.~Cvetic, C.~Dib, C.~S.~Kim and J.~D.~Kim,
%``On lepton flavor violation in tau decays,''
Phys.\ Rev.\ D {\bf 66}, 034008 (2002),
Erratum: [Phys.\ Rev.\ D {\bf 68}, 059901(E) (2003)].

\bibitem{mesons-exp}
J.~P.~Lees {\it et al.} [BaBar Collaboration],
  %``Search for lepton-number violating processes in $B^+ \to h^- l^+ l^+$ decays,''
  Phys.\ Rev.\ D {\bf 85}, 071103 (2012); 
J.~P.~Lees {\it et al.} [BaBar Collaboration],
  %``Search for lepton-number violating BX decays,''
  Phys.\ Rev.\ D {\bf 89}, no. 1, 011102 (2014);
O.~Seon {\it et al.} [BELLE Collaboration],
  %``Search for Lepton-number-violating $B^+ \to D^- l^+ l^{\prime +}$ Decays,''
  Phys.\ Rev.\ D {\bf 84}, 071106 (2011);
R.~Aaij {\it et al.} [LHCb Collaboration],
  %``Searches for Majorana neutrinos in $B^-$ decays,''
  Phys.\ Rev.\ D {\bf 85}, 112004 (2012); 
R.~Aaij {\it et al.} [LHCb Collaboration],
  %``Search for Majorana neutrinos in $B^- \to \pi^+\mu^-\mu^-$ decays,''
  Phys.\ Rev.\ Lett.\  {\bf 112}, no. 13, 131802 (2014);  
J.~Harrison [LHCb and BaBar and Belle Collaborations],
  %``Recent results on searches for heavy Majorana neutrinos,''
  Nucl.\ Part.\ Phys.\ Proc.\  {\bf 260}, 143 (2015).


\bibitem{tau-exp}
Y.~Miyazaki {\it et al.} [Belle Collaboration],
  %``Search for Lepton-Flavor-Violating and Lepton-Number-Violating $\tau \to \ell h h^\prime$ Decay Modes,''
  Phys.\ Lett.\ B {\bf 719}, 346 (2013); 
Y.~Miyazaki {\it et al.} [Belle Collaboration],
  %``Search for Lepton Flavor and Lepton Number Violating tau Decays into a Lepton and Two Charged Mesons,''
  Phys.\ Lett.\ B {\bf 682}, 355 (2010); 
B.~Aubert {\it et al.} [BaBar Collaboration],
  %``Search for lepton-flavor and lepton-number violation in the decay $\tau^- \to \ell^\mp h^\pm h^{\prime -}$,''
  Phys.\ Rev.\ Lett.\  {\bf 95}, 191801 (2005).
%%
\bibitem{Drewes:2015iva} 
  M.~Drewes and B.~Garbrecht,
  %``Combining experimental and cosmological constraints on heavy neutrinos,''
  Nucl.\ Phys.\ B {\bf 921}, 250 (2017).

\bibitem{Canetti:2014dka} 
  L.~Canetti, M.~Drewes and B.~Garbrecht,
  %``Probing leptogenesis with GeV-scale sterile neutrinos at LHCb and Belle II,''
  Phys.\ Rev.\ D {\bf 90}, no. 12, 125005 (2014).

%%
\bibitem{Asaka:2005an} 
T.~Asaka, S.~Blanchet and M.~Shaposhnikov,
  %``The nuMSM, dark matter and neutrino masses,''
  Phys.\ Lett.\ B {\bf 631}, 151 (2005);  
T.~Asaka and M.~Shaposhnikov,
  %``The nuMSM, dark matter and baryon asymmetry of the universe,''
  Phys.\ Lett.\ B {\bf 620}, 17 (2005);
L.~Canetti, M.~Drewes, T.~Frossard and M.~Shaposhnikov,
  %``Dark Matter, Baryogenesis and Neutrino Oscillations from Right Handed Neutrinos,''
  Phys.\ Rev.\ D {\bf 87}, 093006 (2013).

\bibitem{Kayser:1981nw} 
B.~Kayser and R.~E.~Shrock,
  %``Distinguishing Between Dirac and Majorana Neutrinos in Neutral Current Reactions,''
  Phys.\ Lett.\  {\bf 112B}, 137 (1982).

\bibitem{Barranco:2014cda} 
J.~Barranco, D.~Delepine, V.~Gonzalez-Macias, C.~Lujan-Peschard and M.~Napsuciale,
  %``Scattering processes could distinguish Majorana from Dirac neutrinos,''
  Phys.\ Lett.\ B {\bf 739}, 343 (2014);
W.~Rodejohann, X.~J.~Xu and C.~E.~Yaguna,
  %``Distinguishing between Dirac and Majorana neutrinos in the presence of general interactions,''
  JHEP {\bf 1705}, 024 (2017).

\bibitem{Kim:2016bxw} 
C.~S.~Kim and D.~Sahoo,
  %``Deciphering the Majorana nature of neutrino via 'effective' Dalitz plot method,''
  arXiv:1612.00607 [hep-ph].
  
\bibitem{Cvetic:2015naa} 
G.~Cvetic, C.~Dib, C.~S.~Kim and J.~Zamora-Sa\'a,
  %``Probing the Majorana neutrinos and their CP violation in decays of charged
  % scalar mesons $\pi, K, D, D_s, B, B_c$,''
  Symmetry {\bf 7}, 726 (2015). 

\bibitem{PDG2016} 
C.~Patrignani {\it et al.} [Particle Data Group],
  %``Review of Particle Physics,''
  Chin.\ Phys.\ C {\bf 40}, no. 10, 100001 (2016).
  
\bibitem{Cvetic:2014nla} 
G.~Cvetic, C.~S.~Kim and J.~Zamora-Sa\'a, 
  %``CP violation in lepton number violating semihadronic decays of $K,D,D_s,B,B_c$,''
  Phys.\ Rev.\ D {\bf 89}, no. 9, 093012 (2014).

\bibitem{Inami:2016aba} 
  K.~Inami [Belle II Collaboration],
  %``Lepton-flavor-violating $\tau$ decay prospects at SuperKEKB/Belle II,''
  PoS ICHEP {\bf 2016}, 574 (2016).

\bibitem{Aubert:2005tp} 
  B.~Aubert {\it et al.} [BaBar Collaboration],
  %``Search for lepton-flavor and lepton-number violation in the decay $\tau^- \to \ell^\mp h^\pm h^{\prime -}$,''
  Phys.\ Rev.\ Lett.\  {\bf 95}, 191801 (2005).

\bibitem{Miyazaki:2012mx} 
  Y.~Miyazaki {\it et al.} [Belle Collaboration],
  %``Search for Lepton-Flavor-Violating and Lepton-Number-Violating $\tau \to \ell h h^\prime$ Decay Modes,''
  Phys.\ Lett.\ B {\bf 719}, 346 (2013).

\bibitem{Yuan:2017xdp} 
  H.~Yuan, Y.~Jiang, T.~h.~Wang, Q.~Li and G.~L.~Wang,
  %``Lepton Number Violating Four-body Tau Decay,''
  arXiv:1702.04555 [hep-ph].

\end{thebibliography}
\end{document}